\documentclass[prd,showpacs]{revtex4}
\begin{document}
\title{Finite temperature Casimir effect for charged massless scalars in a magnetic field}
\author{Andrea Erdas}
\email{aerdas@loyola.edu}
\author{Kevin P. Seltzer}
\affiliation{Department of Physics, Loyola University Maryland, 4501 North Charles Street,
Baltimore, Maryland 21210, USA}
\begin {abstract} 
The zeta function regularization technique is used to study the finite
temperature Casimir effect
for a charged and massless scalar field confined between parallel plates and satisfying Dirichlet boundary
conditions at the plates. A magnetic field perpendicular to the plates is included. Three equivalent expressions 
for the zeta function are obtained, which are exact to all orders in the magnetic 
field strength, temperature and plate distance. These expressions of the zeta function are used to calculate 
the Helmholtz free energy of the scalar field and the pressure on the plates, in the case of high temperature, small plate distance 
and strong magnetic field. In all cases, simple analytic expressions are obtained for the free energy and pressure
which are accurate and valid for practically all values of temperature, plate distance and magnetic field.
\end {abstract}
\pacs{03.70.+k, 11.10.Wx, 12.20.Ds}
\maketitle
\section{Introduction}
\label{1}
The Casimir effect is a quantum phenomenon where an attractive or repulsive force is observed between 
electrically neutral conducting plates in vacuum, and can be regarded as a quantitative proof
of the quantum fluctuations of the electromagnetic field. Casimir first predicted theoretically the effect,
by calculating the attractive electromagnetic force between two parallel conducting plates \cite{Casimir:1948dh}. 
The repulsive Casimir effect was discovered by Boyer some time later, when he showed that if the electromagnetic field is confined 
inside a perfectly conducting sphere, the wall of the sphere is subject to a repulsive force \cite{Boyer:1968uf}. 
The first experimental evidence of the Casimir force was obtained more than 50 years ago by Sparnaay \cite{Sparnaay:1958wg}
and, since then, many greatly improved experimental observations have been reported. For a comprehensive review 
of these experiments, see the review article and the book by Bordag et al. \cite{Bordag:2001qi,Bordag:2009zz}.

Since Casimir forces have many applications--from nanotubes and nanotechnology \cite{Bellucci:2009jr,Bellucci:2009hh,Golyk:2012nk,
Intravaia:2012iw}, 
to branes and compactified extra dimensions \cite{Poppenhaeger,Edery:2008kd,Pascoal:2007uh,Cheng,Cheng2,
Hertzberg,Hertzberg2,Marachevsky:2007da,Edery,Edery2,Kirsten,Milton,Elizalde,Elizalde2,Flachi,Flachi2,Garriga,
Frank,Linares,Salvio:2012uu,Dupays:2013nm,Jacobs:2012id}, to string theory \cite{Saharian,Hadasz,BezerradeMello:2011nv,Brevik:2012ht}--a 
large effort has gone into studying the Casimir effect
and its generalization to quantum fields other than the electromagnetic field: fermions
 were first considered by Johnson \cite{Johnson:1975} 
in connection with the bag model \cite{Chodos:1974je}, then investigated by many others; for example
\cite{CougoPinto:2001ps,Erdas:2010mz}, bosons and other scalar fields
have also been investigated extensively \cite{Bordag:2001qi}. 

It is well known that Casimir forces
are very sensitive to the boundary conditions of the involved quantum fields on the plates. In the case of scalar fields, the most used boundary conditions
are Dirichlet and Neumann; in the case
of fermion fields or fields with spin in general \cite{Ambjorn:1981xw},
bag boundary conditions are used.
In this work we will use Dirichlet boundary conditions for a scalar field confined between two parallel
plates. 

Scalar fields, with or without charge or mass, appear in many different areas of physics.
The Higgs field is responsible for spontaneous symmetry breaking in the Standard Model and 
is a charged massless scalar before the $SU(2)$ gauge symmetry is broken. Once the symmetry 
is broken, only a neutral massive scalar field remains in the unitary gauge. An ultralight or
massless scalar is the dilaton field that breaks the conformal symmetry of strings in 
superstring theory \cite{Zanzi:2012bf,Kirchbach:2012uz}. Massless scalars called inflatons 
are used to solve the problem of a nonvanishing cosmological constant by causing the 
accelerated expansion of the Universe \cite{Chakravarty:2013eqa,Hindmarsh:2013xg,Boyanovsky:2005yb}.
In condensed matter physics, scalar fields are important to describe spontaneous breaking of
discrete symmetries. The 
Ginzburg-Landau scalar field is associated with type II superconductors and
it was shown that a description of quantum phase excitations in Ginzburg-Landau superconductors
that uses a massless scalar phase field is equivalent to one that uses an antisymmetric 
Kalb-Ramond field \cite{DiGrezia:2004if}. Scalar fields are also
used to explain Landau diamagnetism \cite{Biswas,Brito}, etc. It is well known that the Casimir force between perfectly conducting
parallel plates due to the electromagnetic field is obtained by
multiplying by a factor of 2 the Casimir force due to a massless
scalar field that satisfies Dirichlet boundary conditions on the
plates, where the factor of 2 accounts for the two polarization
states of the photon. Therefore the Casimir force between perfectly conducting parallel plane surfaces
due to a massless, charged scalar field satisfying Dirichlet boundary conditions
on the plates will be the same, apart from a multiplicative factor, as the force
due to a massless, charged vector field satisfying bag boundary conditions on the 
plates. Vector
fields of this type are the $W$ field before symmetry breaking, or the
gluon field in the presence of a chromomagnetic field \cite{Ebert:2001ba,Ebert:2006uh}.

The Casimir effect for charged scalar
fields in a magnetic field has been studied in vacuum \cite{CougoPinto:1998td} and at finite temperature \cite{CougoPinto:1998jg} 
using the Schwinger proper time method to
calculate the effective action, but these authors are only able to obtain the free energy as an infinite sum of 
modified Bessel functions. In this paper we use a different method, the zeta function technique,
to study the Casimir effect for massless scalar fields at finite temperature 
and in the presence of a magnetic field. This method allows us to obtain 
simple analytic forms for the free energy and Casimir pressure, valid for practically all values of the parameters involved.
A similar investigation of the Casimir effect for massive scalar fields at finite temperature 
and in the presence of a magnetic field will be presented elsewhere.

In this paper we will calculate first the Casimir energy for two
parallel plates, and then use it to calculate the Casimir force between the
plates. While the Casimir force between distinct bodies, such as two
parallel plates, is finite, their Casimir energy needs to be
regularized. In the parallel plates case, Casimir effect calculations
must carefully address the issue of regularizing the vacuum energy and therefore
it is best to use the most effective regularization techniques. Many regularization  
techniques are available nowadays, and many of them have been 
applied successfully to the Casimir effect, the cutoff method 
often used in various piston configurations \cite{Cavalcanti:2003tw,Oikonomou:2009zr},
the world-line technique
\cite{Gies:2003cv}, the multiple-scattering method
\cite{Milton:2007wz,Milton:2011zv}, the zeta function technique 
\cite{Elizalde:2007du,Elizalde:2006iu,Elizalde:1988rh}, and others. As we stated above, the choice for this paper is 
the zeta function technique, a
powerful regularization technique used also in the computation of effective actions \cite{dittrich,Erdas:1990gy}. We apply this regularization 
to obtain the free energy and Casimir pressure due to a scalar field confined between two parallel plates, at a distance $a$ from each other.
We assume Dirichlet boundary conditions on the plates and take our system to be in thermal equilibrium 
with a heat reservoir at finite temperature $T$, using the imaginary time 
formalism of finite temperature field theory, which is suitable for a system in thermal equilibrium. 
A uniform magnetic field $\vec B$ is present in the region between the plates and is perpendicular
to the plates.

In Sec. \ref{2}, we obtain three equivalent expressions of the zeta function for this system, exact to all orders in $eB$, $T$ and $a$, where 
$e$ is the scalar field charge. We also obtain simple analytic expressions for the zeta function in the case of 
high temperature ($T\gg a^{-1},  \sqrt{eB}$), small plate distance ($a^{-1}\gg T,  \sqrt{eB}$), and strong magnetic field
($\sqrt{eB}\gg a^{-1},  T$)
In Sec. \ref{3}, we use the zeta function obtained in the previous section, to calculate the Helmholtz free energy 
of the scalar field and the pressure on the plates and obtain simple analytic expressions for these quantities
in the case of high temperature, small plate distance, and strong magnetic field.
A discussion of our results is presented in Sec. \ref{4}. 
\section{ Zeta function evaluation}
\label{2}

Using the imaginary time formalism of finite temperature field theory, we write
the partition function ${\cal Z}$ for a bosonic system in thermal equilibrium at finite temperature $T$
\begin{equation}
{\cal Z}=N\int_{\rm Periodic}\!\!\!\!\!\!\!\!\!\!\!\!\!\!\!
D \phi^\star\,D\phi\, \exp\left(\int_0^\beta d\tau\int d^3x{\cal L}\right),
\label{partition}
\end{equation}
where ${\cal L}$ is the Lagrangian density for the bosonic system, $N$ is a constant and "periodic" means 
that this functional integral is evaluated over field configurations satisfying
\begin{equation}
\phi(x,y,z,\tau)=\phi(x,y,z,\tau+\beta),
\label{antiperiodic}
\end{equation}
for any $\tau$, where $\beta=1/T$ is the periodic length in the Euclidean time axis. In addition to the finite temperature boundary conditions
given by (\ref{antiperiodic}), we impose Dirichlet boundary conditions for scalar bosons between two square plates.
In three-dimensional space with two large parallel plates perpendicular to the $z$ axis
and located at $z=0$ and $z=a$, the Dirichlet boundary conditions constrain the scalar field
to vanish at the plates,
\begin{equation}
\phi(x,y,0,\tau)=\phi(x,y,a,\tau)=0.
\label{bag}
\end{equation}
In the slab region there is also a uniform magnetic field pointing in the $z$ direction, ${\vec B}=(0,0,B)$. The scalar field
has charge $e$ and thus will interact with the magnetic field.

The scalar field Helmholtz free energy $F$ and partition function ${\cal Z}$ are related by 
\begin{equation}
F=-\beta^{-1}\log {\cal Z}.
\label{F}
\end{equation}
A straightforward evaluation of the functional integral (\ref{partition}) yields
\begin{equation}
 \log {\cal Z}=-\log \,\det \left(-D_{\rm E}|{\cal F}_a\right),
\label{logZ}
\end{equation}
where the symbol ${\cal F}_a$ indicates the set of functions which satisfy boundary conditions 
(\ref{antiperiodic}) and (\ref{bag}), and the operator $D_{\rm E}$ is defined as
\begin{equation}
D_{\rm E} = \partial^2_\tau+\partial^2_z-({\vec p} -e{\vec A})^2_\perp,
\label{D}
\end{equation}
where $\vec A$ is the electromagnetic vector potential, the subscript E indicates Euclidean time, and we use
the notation ${\vec p}_\perp=(p_x,p_y,0)$.

The zeta function technique allows us to use the eigenvalues of $D_{\rm E}$ to evaluate $\log {\cal Z}$.
The Dirichlet boundary conditions (\ref{bag}) are satisfied only
if the allowed values for the momentum in the $z$ direction are
\begin{equation}
p_z={\pi\over a}n,
\label{pz}
\end{equation}
where $n \in \{0, 1, 2, 3,...\}$, and therefore
the eigenvalues of $-\partial^2_\tau-\partial^2_z$ whose eigenfunctions satisfy (\ref{antiperiodic}) and (\ref{bag}) are
\begin{equation}
{\pi^2\over a^2}n^2+{4\pi^2\over\beta^2}m^2,
\label{eigenvalues1}
\end{equation}
where $n \in \{0, 1, 2, 3,...\}$ and $m \in \{0, \pm 1, \pm 2, \pm 3,...\}$. The spectrum of the operator 
$({\vec p} -e{\vec A})^2_\perp$ is well known from one-particle quantum mechanics, and its eigenvalues are the Landau levels
\begin{equation}
2eB\left(l+{1\over 2}\right),
\label{eigenvalues2}
\end{equation}
with $l \in \{0, 1, 2, 3,...\}$. Using the eigenvalues (\ref{eigenvalues1}) and (\ref{eigenvalues2}), we construct
the zeta function $\zeta\left(s\right)$, which is given by
\begin{equation}
\zeta(s)=L^2 \left(
{1 \over 2}\right) \sum_{n=-\infty}^\infty \sum_{m=-\infty}^\infty\left(
{eB \over 2\pi}\right) \sum_{l=0}^\infty\left[
{\pi^2\over a^2}n^2+{4\pi^2\over\beta^2}m^2+eB\left(2l+1\right)
\right]^{-s},
\label{zetapm}
\end{equation}
where $L^2$ is the area of the plates and the factor $eB/ 2\pi$ takes into account the degeneracy per unit 
area of the Landau levels. In principle, summation in the index $n$ 
should run from 0 to $\infty$. However, since $n$ appears only squared, 
we run the summation from $-\infty$ to $\infty$ by including a factor of 1/2. 
Note that with this procedure only half the $n=0$ term is taken into account.
This does not affect the physical result because the $n=0$ term contributes to the
Casimir energy a uniform energy density term, and such terms, as we will discuss in Sec. \ref{3},
do not contribute to the Casimir pressure. 

Once we put $\zeta$ in a suitable closed form, using the zeta function technique
we will immediately obtain the partition function
\begin{equation}
\log {\cal Z}=\zeta'(0),
\label{zandzeta}
\end{equation}
and then the free energy using (\ref{F})
\begin{equation}
F=-\beta^{-1}\zeta'(0).
\label{Fandzeta}
\end{equation}
With the help of the following identity
\begin{equation}
z^{-s}={1\over \Gamma(s)}\int_0^\infty dt\, t^{s-1}e^{-zt},
\label{gamma}
\end{equation}
where $\Gamma (s)$ is the Euler gamma function, we rewrite $\zeta(s)$ as
\begin{equation}
\zeta(s)= {L^2 \over 8\pi\Gamma(s)}
\int_0^\infty dt \, t^{s-2} {eBt \over \sinh eBt}\left(\sum_{n=-\infty}^\infty e^{-{\pi^2\over a^2}n^2 t}\right)
\left(\sum_{m=-\infty}^\infty e^{-{4\pi^2\over \beta^2}m^2 t}\right),
\label{zeta2}
\end{equation}
where we also used
\begin{equation}
\sum_{l=0}^\infty e^{-(2l+1)z}={1\over 2 \sinh z}.
\label{sinh}
\end{equation}
It is not possible to evaluate (\ref{zeta2}) in closed form for arbitrary values of $B$, $a$ and $\beta$, but it 
is possible to find simple expressions for $\zeta(s)$  when one or some of $B$, $a$ and $T$ are small or large. From these simple 
expressions of the zeta function,  the free energy will be obtained immediately.

First we evaluate $\zeta(s)$ in the high temperature limit. To do so, we apply the Poisson resummation formula  \cite{Dittrich:1979ux} to 
the $n$ sum in (\ref{zeta2}) and obtain
\begin{equation}
\zeta(s)=a\left[\zeta_{B}(s)+\zeta_{B,a}(s)+\tilde{\zeta}_{B,T}(s) +
\zeta_{B,a,T}(s)\right],
\label{z}
\end{equation}
where
\begin{equation}
{\zeta}_{B}(s)=
{L^2 \over 8\pi^{3/2}\Gamma(s)}
\int_0^\infty dt \, t^{s-5/2} {eBt \over \sinh eBt},
\label{zB}
\end{equation}
\begin{equation}
\tilde{\zeta}_{B,T}(s)=
{L^2 \over 4\pi^{3/2}\Gamma(s)}\sum_{m=1}^\infty 
\int_0^\infty dt \, t^{s-5/2} {eBt \over \sinh eBt} e^{-{4\pi^2m^2t}/\beta^2},
\label{ztBT}
\end{equation}
\begin{equation}
\zeta_{B,a}(s)=
{L^2 \over 4\pi^{3/2}\Gamma(s)}\sum_{n=1}^\infty 
\int_0^\infty dt \, t^{s-5/2} {eBt \over \sinh eBt} e^{-{n^2a^2/t}},
\label{zBa}
\end{equation}
\begin{equation}
\zeta_{B,a,T}(s)=
{L^2 \over 2\pi^{3/2}\Gamma(s)}\sum_{n=1}^\infty \sum_{m=1}^\infty 
\int_0^\infty dt \, t^{s-5/2} {eBt \over \sinh eBt} e^{-({n^2a^2/t} + {4\pi^2m^2t}/\beta^2)}.
\label{zBaT}
\end{equation}
After changing 
the integration variable from $t$ to $t n a\beta /2\pi m$ in  (\ref{zBaT}), we obtain
\begin{equation}
\zeta_{B,a,T}(s)=
{L^2 \over 2\pi^{3/2}\Gamma(s)}\sum_{n=1}^\infty \sum_{m=1}^\infty \left({na\beta\over 2\pi m}\right)^{s-1/2}
\int_0^\infty dt \, t^{s-5/2} {eBt \over \sinh ({eBtna\beta\over 2\pi m})} e^{-2\pi nma(t+1/t)/\beta}.
\label{zBaT2}
\end{equation}
When $2aT \gg1$, only the term with $n=m=1$ contributes significantly to the double sum so, 
using the saddle point method, we evaluate the integral for $eB\ll 4\pi^2T^2$ and obtain
\begin{equation}
\zeta_{B,a,T}(s)=
{L^2 eB \over 2\pi a\Gamma(s)} \left({a\beta\over 2\pi}\right)^{s}
{ e^{-4\pi a/\beta}\over \sinh ({eBa\beta\over 2\pi})}.
\label{zBaT3}
\end{equation}
Next we evaluate (\ref{ztBT}) for $eB\ll 4\pi^2T^2$. In this case,  we can set 
\begin{equation}
{eBt \over\sinh eBt}\approx 1-{1\over 6} (eBt)^2
\label{smallB}
\end{equation}
and, after substituting (\ref{smallB}) into (\ref{ztBT}),  we integrate to find
\begin{equation}
\tilde{\zeta}_{B,T}(s) =
{L^2 \over \Gamma(s)}\left({\beta\over 2\pi}\right)^{2s}
\left[{2\pi^{3/2}\over\beta^3}\Gamma(s-{\scriptstyle{3\over 2}})\zeta_R(2s-3)-
{e^2B^2\beta\over 48\pi^{5/2}}\Gamma(s+{\scriptstyle{1\over 2}})\zeta_R(2s+1)\right],
\label{ztBT2}
\end{equation}
where $\zeta_R$ is the Riemann zeta function of number theory. For
calculating the free energy, we only need to know $\zeta(s)$ for $s\rightarrow 0$. For small $s$ we have
\begin{equation}
z^{2s}\zeta_R(2s-3){\Gamma(s-{\scriptstyle{3\over 2}})\over \Gamma(s)}={\sqrt{\pi}\over 90}s +{\cal O}(s^2),
\label{lim1}
\end{equation}
and
\begin{equation}
z^{2s}\zeta_R(2s+1){\Gamma(s+{\scriptstyle{1\over 2}})\over \Gamma(s)}={\sqrt{\pi}\over 2} 
+\sqrt{\pi}\left(\gamma_E+\ln {z\over 2}\right)s+{\cal O}(s^2),
\label{lim2}
\end{equation}
where $\gamma_E = 0.5772$ is the Euler Mascheroni constant. Substituting (\ref{lim1}) and (\ref{lim2}) into (\ref{ztBT2}), we obtain
\begin{equation}
\tilde{\zeta}_{B,T}(s)=
L^2\left[{\pi^{2}\over 45\beta^3}-
{e^2B^2\beta\over 48\pi^{2}}({1\over 2s}+\gamma_E + \ln{\beta\over 4\pi})\right]s,
\label{ztBT3}
\end{equation}
valid for $eB\ll 4\pi^2T^2$ and small $s$. Notice that Eq. (\ref{ztBT}) is not valid for $s=0$ but, after identifying the presence of Riemann
zeta functions and Euler gamma functions in this equation, and assuming that an analytical 
continuation over the whole complex plane is subtended for these functions,
expressions like Eq. (\ref{ztBT2}) are well behaved for $s\rightarrow 0$. The same is true for Eqs.
(\ref{zB}) and (\ref{zBa}): they are not valid for $s=0$ but, once Riemann
zeta functions and Euler gamma functions are identified inside these equations and
analytic continuation is subtended, they will become well behaved for $s\rightarrow 0$.

In the high temperature limit, both $eBa^2\ll1$ and $eBa^2\gg1$ are possible, and therefore we need to evaluate
 (\ref{zBa}) for both scenarios. When $eBa^2\ll1$ we use (\ref{smallB}) in (\ref{zBa}), integrate, and find
 \begin{equation}
{\zeta}_{B,a}(s) =
{L^2 a^{2s}\over 4\pi^{3/2}\Gamma(s)}
\left[{1\over a^{3}}\Gamma({\scriptstyle{3\over 2}}-s)\zeta_R(3-2s)-
{e^2B^2 a\over 6}\Gamma(-{\scriptstyle{1\over 2}}-s)\zeta_R(-1-2s)\right],
\label{zBa2}
\end{equation}
which, for small $s$, becomes
 \begin{equation}
{\zeta}_{B,a}(s) =
{L^2 \over 8\pi}
\left[{\zeta_R(3)\over a^{3}}-
{e^2B^2 a\over 18}\right]s,
\label{zBa3}
\end{equation}
where $\zeta_R(3) = 1.2021$. When $eBa^2\gg1$ we use 
\begin{equation}
{1\over\sinh eBt}\approx 2e^{-eBt}
\label{largeB}
\end{equation}
in (\ref{zBa}), change
the integration variable from $t$ to ${n a \over \sqrt{eB}}t$, and find
\begin{equation}
\zeta_{B,a}(s)=
{L^2 {eB}\over 2\pi^{3/2}\Gamma(s)}\sum_{n=1}^\infty \left({n a \over \sqrt{eB}}\right)^{s-1/2}
\int_0^\infty dt \, t^{s-3/2}  e^{-{\sqrt{eB}na(t+1/t)}}.
\label{zBa4}
\end{equation}
Only the term with $n=1$ contributes significantly to the sum when $eBa^2\gg1$ and, 
using the saddle point method, we evaluate the integral and find
\begin{equation}
\zeta_{B,a}(s)=
{L^2 {eB}\over 2\pi a\Gamma(s)} \left({ a \over \sqrt{eB}}\right)^{s}
 e^{-2{\sqrt{eB}a}}.
\label{zBa5}
\end{equation}

Finally, we calculate $\zeta_B(s)$, the only piece of the zeta function that can be evaluated exactly and,
after integrating, we find
\begin{equation}
{\zeta}_{B}(s)=
{L^2 (eB)^{3/2-s}\over 4\pi^{3/2}\Gamma(s)}
(1-2^{1/2-s})\Gamma(s-{\scriptstyle{1\over 2}})\zeta_R (s-{\scriptstyle{1\over 2}}),
\label{zB2}
\end{equation}
which, for small $s$, becomes
\begin{equation}
{\zeta}_{B}(s)=
{L^2 (eB)^{3/2}\over 2\pi}(\sqrt{2}-1)
\zeta_R (-{\scriptstyle{1\over 2}})s,
\label{zB3}
\end{equation}
where $\zeta_R (-{\scriptstyle{1\over 2}})=-0.2079$.

By adding (\ref{zBaT3}), (\ref{ztBT3}), (\ref{zBa3}) and (\ref{zB3}), we find $\zeta(s)$ in the high temperature and
very weak field limit, $2T\gg a^{-1}$, $2T\gg \sqrt{eB}/\pi$, and $eB\ll a^{-2}$,
\begin{equation}
\zeta(s)=L^2\left[{\pi^{2}a\over 45\beta^3}+
{ (eB)^{3/2}a\over 2\pi}(\sqrt{2}-1)\zeta_R (-{\scriptstyle{1\over 2}})
+ {\zeta_R(3)\over 8\pi a^{2}}+
{eB \over 2\pi} { e^{-4\pi a/\beta}\over \sinh ({eBa\beta\over 2\pi})}
-{e^2B^2 a^2\over 144\pi}-
{e^2B^2\beta a\over 48\pi^{2}}({1\over 2s}+\gamma_E + \ln{\beta\over 4\pi})\right]s,
\label{z3}
\end{equation}
where we took the small $s$ limit.
By adding (\ref{zBaT3}), (\ref{ztBT3}), (\ref{zBa5}) and (\ref{zB3}), we find $\zeta(s)$ in the high temperature and
very large plate distance limit, $2T\gg a^{-1}$, $2T\gg \sqrt{eB}/\pi$, and $eB\gg a^{-2}$,
\begin{equation}
\zeta(s)=
L^2\left[{\pi^{2}a\over 45\beta^3}+
{ (eB)^{3/2}a\over 2\pi}(\sqrt{2}-1)\zeta_R (-{\scriptstyle{1\over 2}})+
{eB \over 2\pi} { e^{-4\pi a/\beta}\over \sinh ({eBa\beta\over 2\pi})}+
{eB \over 2\pi} e^{-2{\sqrt{eB}a}}-
{e^2B^2\beta a\over 48\pi^{2}}({1\over 2s}+\gamma_E + \ln{\beta\over 4\pi})\right]s,
\label{z4}
\end{equation}
where we also took the small $s$ limit.

Next we evaluate $\zeta(s)$ in the limit of small plate distance and apply the Poisson resummation formula to the $m$ sum in 
(\ref{zeta2}) to obtain
\begin{equation}
\zeta(s)={\beta\over 2}\left[\zeta_{B}(s)+\tilde{\zeta}_{B,a}(s)+\zeta_{B,T}(s) +
\tilde{\zeta}_{B,a,T}(s)\right],
\label{z2}
\end{equation}
where $\zeta_B(s)$ is the same as in (\ref{zB}), and
\begin{equation}
\tilde{\zeta}_{B,a}(s)=
{L^2 \over 4\pi^{3/2}\Gamma(s)}\sum_{n=1}^\infty 
\int_0^\infty dt \, t^{s-5/2} {eBt \over \sinh eBt} e^{-{\pi^2n^2t}/a^2},
\label{ztBa}
\end{equation}
\begin{equation}
\zeta_{B,T}(s)=
{L^2 \over 4\pi^{3/2}\Gamma(s)}\sum_{m=1}^\infty 
\int_0^\infty dt \, t^{s-5/2} {eBt \over \sinh eBt} e^{-{m^2\beta^2/4t}},
\label{zBT}
\end{equation}
\begin{equation}
\tilde{\zeta}_{B,a,T}(s)=
{L^2 \over 2\pi^{3/2}\Gamma(s)}\sum_{n=1}^\infty \sum_{m=1}^\infty 
\int_0^\infty dt \, t^{s-5/2} {eBt \over \sinh eBt} e^{-(\pi^2n^2t/a^2+m^2\beta^2/4t)}.
\label{ztBaT}
\end{equation}
It is evident from (\ref{z2}) - (\ref{ztBaT}) that $\zeta(s)$, in the limits $2aT\ll 1$ and $eB\ll\pi^2a^{-2}$,
is obtained from (\ref{z}) - (\ref{zBaT})  by replacing $a$ with $\beta/2$ and $\beta$ with $2a$. For $eB\left({\beta\over 2}\right)^2\ll 1$ 
and small $s$, we find
\begin{equation}
\zeta(s)=L^2\left[
{\pi^{2}\beta\over 720a^3}+
{ (eB)^{3/2}\beta\over 4\pi}(\sqrt{2}-1)\zeta_R (-{\scriptstyle{1\over 2}})+
{\zeta_R(3)\over 2\pi \beta^{2}}+
{eB \over 2\pi} { e^{-\pi \beta/a}\over \sinh ({eBa\beta\over 2\pi})}
-{e^2B^2 \beta^2\over 576\pi}-
{e^2B^2\beta a\over 48\pi^{2}}({1\over 2s}+\gamma_E + \ln{a\over 2\pi})\right]s,
\label{z5}
\end{equation}
and for  $eB\left({\beta\over 2}\right)^2\gg 1$ and small $s$, we find
\begin{equation}
\zeta(s)=L^2\left[
{\pi^{2}\beta\over 720a^3}+
{ (eB)^{3/2}\beta\over 4\pi}(\sqrt{2}-1)\zeta_R (-{\scriptstyle{1\over 2}})+
{eB \over 2\pi} { e^{-\pi \beta/a}\over \sinh ({eBa\beta\over 2\pi})}+
{eB \over 2\pi} e^{-{\sqrt{eB}\beta}}-
{e^2B^2\beta a\over 48\pi^{2}}({1\over 2s}+\gamma_E + \ln{a\over 2\pi})\right]s.
\label{z6}
\end{equation}

Last, we evaluate $\zeta(s)$ in the strong magnetic field limit, $eB\gg \left({\beta\over 2}\right)^{-2}$ and $eB\gg a^{-2}$. 
Under these conditions, after applying the Poisson resummation formula to both the $n$ and $m$ sums in (\ref{zeta2}), we find
\begin{equation}
\zeta(s)=a\beta[ \zeta_{W}(s)+\tilde\zeta(s)]
\label{z7}
\end{equation}
where
\begin{equation}
\zeta_{W}(s)= {L^2 \over 16\pi^2\Gamma(s)}
\int_0^\infty dt \, t^{s-3} {eBt \over \sinh eBt},
\label{zW}
\end{equation}
is the zeta function of the  one-loop vacuum effective Lagrangian 
for massless scalar QED first calculated by Weisskopf, and 
\begin{equation}
\tilde\zeta(s)= {L^2 \over 16\pi^2\Gamma(s)}
\int_0^\infty dt \, t^{s-3} {eBt \over \sinh eBt}\left(\sum_{n,m=-\infty}^\infty e^{-{a^2}n^2 /t} e^{-{ \beta^2}m^2/4 t}-1\right).
\label{z8}
\end{equation}
The integral in (\ref{zW}) can be evaluated exactly, and we find
\begin{equation}
{\zeta}_{W}(s)=
{L^2 (eB)^{2-s}\over 8\pi^{2}\Gamma(s)}
(1-2^{1-s})\Gamma(s-1)\zeta_R (s-1),
\label{zW2}
\end{equation}
which, for small $s$, becomes
\begin{equation}
{\zeta}_{W}(s)=
{L^2 e^2B^{2}\over 96\pi^2}
\left(\ln eB -\ln 3-{1\over 2} -{1\over s}\right)s,
\label{zW3}
\end{equation}
where we used the interesting numerical fact \cite{Erdas:2010yq}
\begin{equation}
\gamma_E+\ln \pi -{6\over \pi^2}\zeta'(2)\simeq \ln 6 +{1\over 2}.
\label{num}
\end{equation}

We evaluate $\tilde\zeta(s)$ by using (\ref{largeB}), which is valid in the strong magnetic field limit; we then change the integration variable from $t$ to
$\sqrt{n^2 a^2+m^2\beta^2/4 \over {eB}}t$, to find
\begin{equation}
\tilde\zeta(s)= {L^2 eB\over 8\pi^2\Gamma(s)} \sum_{n,m=-\infty}^\infty
\left({n^2 a^2+m^2\beta^2/4 \over {eB}}\right)^{{s-1}\over 2}\int_0^\infty dt \, t^{s-2} 
e^{-(t+1/t)\sqrt{eB}{\sqrt{n^2 a^2+m^2\beta^2/4 }}},
\label{z9}
\end{equation}
where the term with $m=n=0$ is excluded and only terms with $n=0,\pm 1$ and $m=0,\pm 1$ contribute significantly to the double sum. 
We integrate using the saddle point method and, for small $s$, obtain
\begin{equation}
\tilde\zeta(s)= {L^2 (eB)^{5/4}  \over 2\pi^{3/2} }\left[{e^{-2a\sqrt{eB}}\over 2a^{3/2}}+{\sqrt{2}e^{-\beta\sqrt{eB}}\over \beta^{3/2}}+
{e^{-2\sqrt{eB}\sqrt{a^2+\beta^2/4 }}\over \left({a^2+\beta^2/4 }\right)^{3/4}}\right]s.
\label{z10}
\end{equation}
Adding (\ref{zW3}) to (\ref{z10}) we find the zeta function in the strong magnetic field  limit,
$eB\gg a^{-2}$ and $eB\gg \left({\beta\over 2}\right)^{-2}$, and small $s$
\begin{equation}
\zeta(s)=L^2a\beta\left[{e^2B^{2}\over 96\pi^2}
\left(\ln eB -\ln 3-{1\over 2} -{1\over s}\right)+
{(eB)^{5/4}  \over 2\pi^{3/2} }\left({e^{-2a\sqrt{eB}}\over 2a^{3/2}}+{\sqrt{2}e^{-\beta\sqrt{eB}}\over \beta^{3/2}}+
{e^{-2\sqrt{eB}\sqrt{a^2+\beta^2/4 }}\over \left({a^2+\beta^2/4 }\right)^{3/4}}\right)
\right]s.
\label{z11}
\end{equation}
\section{ Free energy and Casimir pressure}
\label{3}
The derivative of the zeta function is obtained easily by taking advantage of the useful
fact  that, for a well-behaved $G(s)$, the derivative of $G(s)/\Gamma(s)$ at $s=0$ is simply $G(0)$ and
therefore, using (\ref{Fandzeta}) and (\ref{zeta2}), we find the free energy
\begin{equation}
F=-{L^2 \over 8\pi\beta}
\int_0^\infty dt \, t^{-2} {eBt \over \sinh eBt}\left(\sum_{n=-\infty}^\infty e^{-{\pi^2\over a^2}n^2 t}\right)
\left(\sum_{m=-\infty}^\infty e^{-{4\pi^2\over \beta^2}m^2 t}\right).
\label{F2}
\end{equation}
Using our results for the zeta function (\ref{z}), (\ref{z2}),  and (\ref{z7}), we are able to obtain three other expressions of 
the free energy, all equivalent to  (\ref{F2}), 
\begin{equation}
F=-
{L^2a \over 8\pi^{3/2}\beta}
\int_0^\infty dt \, t^{-5/2} {eBt \over \sinh eBt}\left(\sum_{n=-\infty}^\infty e^{-{n^2a^2\over t}}\right)
\left(\sum_{m=-\infty}^\infty e^{-{4\pi^2\over \beta^2}m^2 t}\right),
\label{F2a}
\end{equation}
best suited for high temperature expansion ($2Ta\gg 1$ and $2T\gg \sqrt{eB}/\pi$),
\begin{equation}
F=-
{L^2 \over 16\pi^{3/2}}
\int_0^\infty dt \, t^{-5/2} {eBt \over \sinh eBt}\left(\sum_{n=-\infty}^\infty e^{-{\pi^2\over a^2}n^2 t}\right)
\left(\sum_{m=-\infty}^\infty e^{-{m^2 \beta^2\over 4t} }\right),
\label{F2b}
\end{equation}
best suited for small plate distance expansion ($2Ta\ll 1$ and $a^{-1}\gg \sqrt{eB}/\pi$), and
\begin{equation}
F=-
{L^2a \over 16\pi^{2}}
\int_0^\infty dt \, t^{-3} {eBt \over \sinh eBt}\left(\sum_{n=-\infty}^\infty e^{-{n^2a^2\over t}}\right)
\left(\sum_{m=-\infty}^\infty e^{-{m^2 \beta^2\over 4t} }\right),
\label{F2c}
\end{equation}
best suited for strong magnetic field expansion. The last equation has been obtained by other authors
\cite{CougoPinto:1998jg}, who used (\ref{F2c}) to write the free energy as an infinite sum of
modified Bessel functions.

It is not possible to evaluate (\ref{F2}) - (\ref{F2c}) in closed form for arbitrary values of $B$, $a$ 
and $\beta$ but, using our results from Sec. \ref{2}, we found simple analytic expressions for the free energy when one or some of 
those three quantities are small or large. 
To calculate the free energy in the high temperature limit, $2T\gg a^{-1}$ and $2T\gg \sqrt{eB}/\pi$, we use 
(\ref{z3}) and (\ref{z4}) to find  
\begin{equation}
F=-L^2\left[
{\pi^{2}a\over 45\beta^4}+
{ (eB)^{3/2}a\over 2\pi\beta}(\sqrt{2}-1)\zeta_R (-{\scriptstyle{1\over 2}})
+{\zeta_R(3)\over 8\pi \beta a^{2}}+
{eB \over 2\pi\beta} { e^{-4\pi a/\beta}\over \sinh ({eBa\beta\over 2\pi})}
-{e^2B^2 a^2\over 144\pi\beta}-
{e^2B^2 a\over 48\pi^{2}}(\gamma_E + \ln{\beta\over 4\pi})\right],
\label{F3}
\end{equation}
valid for $eB\ll a^{-2}$, and
\begin{equation}
F=-L^2\left[
{\pi^{2}a\over 45\beta^4}+
{ (eB)^{3/2}a\over 2\pi\beta}(\sqrt{2}-1)\zeta_R (-{\scriptstyle{1\over 2}})
+{eB \over 2\pi\beta} { e^{-4\pi a/\beta}\over \sinh ({eBa\beta\over 2\pi})}+
{eB \over 2\pi\beta} e^{-2{\sqrt{eB}a}}-
{e^2B^2 a\over 48\pi^{2}}(\gamma_E + \ln{\beta\over 4\pi})\right],
\label{F4}
\end{equation}
valid for $eB\gg a^{-2}$. Notice that in (\ref{F3}) and (\ref{F4}) the dominant term is the Stefan-Boltzmann term $-{\pi^{2}\over 45}VT^4$, 
where $V=L^2a$ is the volume of the slab. Terms with a linear dependence on the plate distance,
such as this one, are 
proportional to the volume of the slab and represent a uniform energy density. If the same magnetic
field is present outside the slab and the medium outside the slab is also at temperature $T$, such terms do not
contribute to the Casimir pressure. If there is vacuum outside the slab, i.e. no magnetic field and zero temperature, uniform energy
density terms contribute a constant pressure which is very easily calculated. In this paper we assume that the same
magnetic field is present inside and outside the slab, and that the medium outside the slab is at the same temperature as the one inside the
slab, so we neglect contributions to the Casimir pressure from uniform energy density terms.

The pressure $P$ on the plates is given by
\begin{equation}
P=-{1\over L^2}{\partial F\over\partial a},
\label{P1}
\end{equation}
and therefore, for $2T\gg a^{-1}\gg \sqrt{eB}/\pi$, we find
\begin{equation}
P=- {\zeta_R(3)\over 4\pi \beta a^{3}}-
{2eB \over \beta^2} { e^{-4\pi a/\beta}\over \sinh ({eBa\beta\over 2\pi})}-
{ e^2B^2e^{-4\pi a/\beta} \over 4\pi^2}{ \coth ({eBa\beta\over 2\pi})\over \sinh ({eBa\beta\over 2\pi})}
-{e^2B^2 a\over 72\pi\beta},
\label{P2}
\end{equation}
and
\begin{equation}
P=- {2eB \over \beta^2} { e^{-4\pi a/\beta}\over \sinh ({eBa\beta\over 2\pi})}-
{ e^2B^2e^{-4\pi a/\beta} \over 4\pi^2}{ \coth ({eBa\beta\over 2\pi})\over \sinh ({eBa\beta\over 2\pi})}
-{(eB)^{3/2} \over \pi\beta} e^{-2{\sqrt{eB}a}},
\label{P3}
\end{equation}
for $2T\gg \sqrt{eB}/\pi\gg a^{-1}$.
Since the third term in (\ref{P2}) is negligible when compared to the other ones in (\ref{P2}) and (\ref{P3}), we write
the high temperature Casimir pressure as
\begin{equation}
P=- {\zeta_R(3)\over 4\pi \beta a^{3}}-
{2eB \over \beta^2} { e^{-4\pi a/\beta}\over \sinh ({eBa\beta\over 2\pi})}-{e^2B^2 a\over 72\pi\beta},
\label{P4}
\end{equation}
for $2T\gg a^{-1}\gg \sqrt{eB}/\pi$, and
\begin{equation}
P= - {2eB \over \beta^2} { e^{-4\pi a/\beta}\over \sinh ({eBa\beta\over 2\pi})}
-{(eB)^{3/2} \over \pi\beta} e^{-2{\sqrt{eB}a}},
\label{P5}
\end{equation}
for $2T\gg \sqrt{eB}/\pi\gg a^{-1}$.

Next we obtain the free energy in the small plate distance limit, using  (\ref{z5}) and (\ref{z6}), to find
\begin{equation}
F=-L^2\left[
{\pi^{2}\over 720a^3}+
{ (eB)^{3/2}\over 4\pi}(\sqrt{2}-1)\zeta_R (-{\scriptstyle{1\over 2}})+
{\zeta_R(3)\over 2\pi \beta^{3}}+
{eB \over 2\pi\beta} { e^{-\pi \beta/a}\over \sinh ({eBa\beta\over 2\pi})}
-{e^2B^2 \beta\over 576\pi}-
{e^2B^2 a\over 48\pi^{2}}(\gamma_E + \ln{a\over 2\pi})\right]
\label{F5}
\end{equation}
for  $a^{-1}\gg 2T\gg \sqrt{eB}/\pi$, and
\begin{equation}
F=-L^2\left[
{\pi^{2}\over 720a^3}+
{ (eB)^{3/2}\over 4\pi}(\sqrt{2}-1)\zeta_R (-{\scriptstyle{1\over 2}})+
{eB \over 2\pi\beta} { e^{-\pi \beta/a}\over \sinh ({eBa\beta\over 2\pi})}+
{eB \over 2\pi\beta} e^{-{\sqrt{eB}\beta}}-
{e^2B^2 a\over 48\pi^{2}}(\gamma_E + \ln{a\over 2\pi})\right]
\label{F6}
\end{equation}
for  $a^{-1}\gg \sqrt{eB}/\pi\gg 2T$. The dominant term here is $-{\pi^{2}\over 720}{L^2\over a^3}$, which is the familiar vacuum Casimir energy 
for a complex scalar field and for the photon field \cite{Casimir:1948dh}.
The Casimir pressure for small plate distance is
\begin{equation}
P=-
{\pi^{2}\over 240 a^4}+
{eB \over 2a^2} { e^{-\pi \beta/a}\over \sinh ({eBa\beta\over 2\pi})}-
{ e^2B^2e^{-\pi \beta/a} \over 4\pi^2}{ \coth ({eBa\beta\over 2\pi})\over \sinh ({eBa\beta\over 2\pi})}
-{e^2B^2 \over 48\pi^{2}}( \ln{a\over 2\pi}+1)
\label{P6}
\end{equation}
in the case  of very weak magnetic field ($a^{-1}\gg 2T\gg \sqrt{eB}/\pi$), and  it is identical in the case of
 very low temperature ($ a^{-1}\gg \sqrt{eB}/\pi\gg 2T$).
 Since the third term in (\ref{P6}) is much smaller than the other ones, we can neglect it and write the
 pressure in the small plate distance limit as
\begin{equation}
P=-
{\pi^{2}\over 240 a^4}+
{eB \over 2a^2} { e^{-\pi \beta/a}\over \sinh ({eBa\beta\over 2\pi})}
-{e^2B^2 \over 48\pi^{2}}( \ln{a\over 2\pi}+1).
\label{P7}
\end{equation}

Finally, for strong magnetic field the free energy is found using (\ref{z11})
\begin{equation}
F=-L^2a\left[{e^2B^{2}\over 96\pi^2}
\left(\ln eB -\ln 3-{1\over 2} \right)+
{(eB)^{5/4}  \over 2\pi^{3/2} }\left({e^{-2a\sqrt{eB}}\over 2a^{3/2}}+{\sqrt{2}e^{-\beta\sqrt{eB}}\over \beta^{3/2}}+
{e^{-2\sqrt{eB}\sqrt{a^2+\beta^2/4 }}\over \left({a^2+\beta^2/4 }\right)^{3/4}}\right)
\right],
\label{F7}
\end{equation}
where the dominant term is the one-loop vacuum effective potential for massless scalar QED \cite{Erdas:2010yq}, and it is
proportional to the volume of the slab, as expected. The effective potential is a uniform energy density term and 
therefore, under our assumptions, does not contribute to the Casimir pressure.
The pressure, for $eB\gg \left({\beta\over 2}\right)^{-2}$ and $eB\gg a^{-2}$, is given by
\begin{equation}
P=-{(eB)^{5/4}  \over 2\pi^{3/2}\sqrt{a} }{e^{-2a\sqrt{eB}} }\left(\sqrt{eB}+{1\over 4 a }\right)
+{(eB)^{5/4}  \over 2\pi^{3/2} }{e^{-2\sqrt{eB}\sqrt{a^2+\beta^2/4 }}\over \left({a^2+\beta^2/4 }\right)^{3/4}}
\left(1-{2a^2\sqrt{eB}\over  \sqrt{a^2+\beta^2/4 }}-{3\over 2}{a^2\over a^2+\beta^2/4}\right),
\label{P8}
\end{equation}
and, neglecting the smaller terms, we obtain
\begin{equation}
P=-{(eB)^{7/4}  \over \pi^{3/2} }\left(
{{e^{-2a\sqrt{eB}} }\over 2\sqrt{a} }+
{a^2e^{-2\sqrt{eB}\sqrt{a^2+\beta^2/4 }}\over  \left({a^2+\beta^2/4 }\right)^{5/4}}\right).
\label{P9}
\end{equation}
\section{Discussion and conclusions}
\label{4}
In this paper we used the zeta function regularization technique to study the finite
temperature Casimir effect
of a massless charged scalar field confined between parallel plates and in the 
presence of a magnetic field perpendicular to the plates. We have obtained three expressions 
 for the zeta function (\ref{z}), (\ref{z2}),  and (\ref{z7}),  which are exact to all orders in the magnetic 
field strength $B$, plate distance $a$ and inverse temperature $\beta$, and we have used them to derive expressions for the Helmholtz free
energy and for the Casimir pressure on the plates in the case of high temperature
($4T^2\gg a^{-2},eB/\pi^2 $), small plate distance ($a^{-2}\gg 4T^{2},eB/\pi^2$) and strong magnetic field
($eB\gg a^{-2}, 4T^2$).

We have been able to numerically evaluate the free energy with very high precision, using the three exact expressions 
(\ref{F2a}), (\ref{F2b}),  and (\ref{F2c}),  and we compared the values of the free energy obtained from
our simple analytic expressions to the exact numerical values. In the high temperature case we found that, for $2aT=4$,
Eq. (\ref{F3}) is within $0.7$ percent of the exact value of the free energy in the range $0\le eBa^2 \le 1$, while 
Eq. (\ref{F4}) is within $0.7$ percent of the exact value of the free energy in the range $1\le eBa^2 \le \infty$. 
For $2aT=10$, Eq. (\ref{F3}) is within $0.05$ percent of the exact value of the free energy in the range $0\le eBa^2 \le 1$, while 
Eq. (\ref{F4}) is within $0.05$ percent of the exact value of the free energy in the range $1\le eBa^2 \le \infty$, showing a very 
rapid decrease of the small discrepancy between Eqs. (\ref{F3}), (\ref{F4}) and the exact values of the free energy.
We summarize this finding by writing the free energy per unit area in the high temperature limit as
\begin{equation}
{F\over L^2}=\cases{
-{\pi^{2}a\over 45\beta^4}-
{ (\sqrt{2}-1)\zeta_R (-{\scriptstyle{1\over 2}})(eB)^{3/2}a\over 2\pi\beta}
-{eB \over 2\pi\beta} { e^{-4\pi a/\beta}\over \sinh ({eBa\beta\over 2\pi})}
-{\zeta_R(3)\over 8\pi \beta a^{2}}+{e^2B^2 a^2\over 144\pi\beta}+
{e^2B^2 a\over 48\pi^{2}}(\gamma_E + \ln{\beta\over 4\pi})
& \text{for $0\le eBa^2 \le 1$ ;}
\cr
-{\pi^{2}a\over 45\beta^4}-
{ (\sqrt{2}-1)\zeta_R (-{\scriptstyle{1\over 2}})(eB)^{3/2}a\over 2\pi\beta}
-{eB \over 2\pi\beta} { e^{-4\pi a/\beta}\over \sinh ({eBa\beta\over 2\pi})}-
{eB \over 2\pi\beta} e^{-2{\sqrt{eB}a}}+
{e^2B^2 a\over 48\pi^{2}}(\gamma_E + \ln{\beta\over 4\pi})
& 
\text{for $1\le eBa^2 < \infty$.}
\cr}
\label{F8}
\end{equation}
Eq. (\ref{F8}) is a simple analytic expression 
of $F$ in the high temperature limit, valid for all values of 
the magnetic field $B$ and the plate distance $a$, and with a discrepancy of no more than $0.7$ percent from 
the exact value of $F$ for $2aT\ge4$. A similarly accurate expression of the Casimir pressure $P$, valid for $2aT\ge4$ and all values of $a$ 
and $B$, is obtained immediately from (\ref{F8}), since $P=-{1\over L^2}{\partial F\over\partial a}$. 
To roughly indicate in what regimes of temperature, magnetic field, and plate separation
Eq. (\ref{F8}) holds, we give two numerical examples for the high temperature limit, 
one with $T=10^4$K and the other with $T=10^6$K, and we take the charge $e$ of the scalar field to
equal the elementary charge.
For $T=10^4$K, corresponding
to $8.62\times 10^{-1}$eV, Eq. (\ref{F8}) is valid for $a^{-1}\le 4.31\times 10^{-1}$eV in natural
units, corresponding to $a \ge 4.57\times 10^{-7}$m in SI units. Given a value of $a$
within this range, for example $a=10^{-5}$m, the top part of (\ref{F8}) should be 
used when $eB\le 3.88\times 10^{-4}$eV$^2$ in natural 
units, which corresponds to $B\le 6.55\times 10^{-2}$G in cgs units, while the bottom part should be used when
$B\ge 6.55\times 10^{-2}$G.
For $T=10^6$K, Eq. (\ref{F8}) is valid when $a \ge 4.57\times 10^{-9}$m. For a value of $a$
within this range, for example $a=10^{-6}$m, the top part of (\ref{F8})
is valid when $B\le 6.55$G, and the bottom part is valid when
$B\ge 6.55$G.

In the small plate distance case we found that, for $2aT={1\over 4}$,
Eq. (\ref{F5}) is within $0.7$ percent of the exact value of the free energy in the range $0\le eB\left({\beta\over 2}\right)^2 \le 1$, while 
Eq. (\ref{F6}) is within $0.7$ percent of the exact value of the free energy 
in the range $1\le eB\left({\beta\over 2}\right)^2 \le \infty$.
For $2aT={1\over 10}$,
Eq. (\ref{F5}) is within $0.05$ percent of the exact value of the free energy in the range $0\le eB\left({\beta\over 2}\right)^2 \le 1$, while 
Eq. (\ref{F6}) is within $0.05$ percent of the exact value of the free energy in the range $1\le eB\left({\beta\over 2}\right)^2 \le \infty$, showing again
a very rapid decrease of the small discrepancy between our analytical expressions and the exact values of the free energy.
We summarize the small plate distance limit by writing the free energy per unit area as
\begin{equation}
{F\over L^2}=\cases{
-{\pi^{2}\over 720a^3}-
{ (\sqrt{2}-1)\zeta_R (-{\scriptstyle{1\over 2}})(eB)^{3/2}\over 4\pi}-
{eB \over 2\pi\beta} { e^{-\pi \beta/a}\over \sinh ({eBa\beta\over 2\pi})}-
{\zeta_R(3)\over 2\pi \beta^{3}}
+{e^2B^2 \beta\over 576\pi}+
{e^2B^2 a\over 48\pi^{2}}(\gamma_E + \ln{a\over 2\pi})
& \text{for $0\le eB\left({\beta\over 2}\right)^2 \le 1$ ;}
\cr
-{\pi^{2}\over 720a^3}-
{ (\sqrt{2}-1)\zeta_R (-{\scriptstyle{1\over 2}})(eB)^{3/2}\over 4\pi}-
{eB \over 2\pi\beta} { e^{-\pi \beta/a}\over \sinh ({eBa\beta\over 2\pi})}-
{eB \over 2\pi\beta} e^{-{\sqrt{eB}\beta}}+
{e^2B^2 a\over 48\pi^{2}}(\gamma_E + \ln{a\over 2\pi})
& 
\text{for $1\le eB\left({\beta\over 2}\right)^2 < \infty$,}
\cr}
\label{F9}
\end{equation}
a simple analytic expression of $F$, valid for all values of 
$B$ and $T$, and with a discrepancy of no more than $0.7$ percent from 
the exact value of $F$ for $2aT\le{1\over 4}$. The pressure in the case of small plate distance is obtained immediately from
(\ref{F9}) for $2aT\le{1\over 4}$ and all values of $B$ and $T$. 
We now give two numerical examples for the small plate distance case, 
one with $T=100$K and the other with $T=300$K.
For $T=100$K, Eq. (\ref{F9}) is valid for $a \le 2.86\times 10^{-4}$m. The top part of (\ref{F9}) 
should be used when $B\le 5.01\times 10^{-2}$G, while the bottom part should be used when
$B\ge 5.01\times 10^{-2}$G.
For $T=300$K, Eq. (\ref{F9}) is valid for $a \le 9.53\times 10^{-5}$m. The top part of (\ref{F9}) 
should be used when 
$B\le 4.50\times 10^{-1}$G, while the bottom part should be used when
$B\ge 4.50\times 10^{-1}$G.
Notice that, if we set $T=0$ in (\ref{F9}), we obtain the Casimir energy $E_C$ for a massless and charged
scalar field in a magnetic field,
\begin{equation}
{E_C \over L^2}=
-{\pi^{2}\over 720a^3}-
{ (\sqrt{2}-1)\zeta_R (-{\scriptstyle{1\over 2}})(eB)^{3/2}\over 4\pi}+
{e^2B^2 a\over 48\pi^{2}}(\gamma_E + \ln{a\over 2\pi}),
\label{EC1}
\end{equation}
where we see that the magnetic field, as it grows, inhibits the Casimir energy of the scalar field \cite{CougoPinto:1998td}. 
Our result, a simple analytic expression for $E_C$, is more explicit than that of \cite{CougoPinto:1998td} where the magnetic field correction
to the Casimir energy is presented as an infinite sum of integrals.

In the case of strong magnetic field, the free energy shown in Eq. (\ref{F7}) is valid for all values of $a$ and $T$, and so is the
pressure shown in Eq. (\ref{P9}). If we set $T=0$ in (\ref{F7}), we can neglect the effective potential which is a uniform energy density term, 
and obtain $E_C$ in the strong magnetic field case,
\begin{equation}
{E_C \over L^2}=-{1\over4\pi^{3/2} }
{(eB)^{5/4} e^{-2a\sqrt{eB}} \over \sqrt{a} }
\label{EC2}
\end{equation}
which agrees with \cite{CougoPinto:1998td} on the dependence of $E_C$ from $a$ and $B$, but is in disagreement
for the overall sign, since we obtain a negative value for $E_C$, not a positive one. We also obtain the 
numerical constant present in $E_C$, while the authors of Ref. \cite{CougoPinto:1998td} did not. 

We conclude with
a brief discussion of how observable this effect is. For a plate distance $a=1$  $\mu$m and a magnetic field $B=100$ G, 
$eB$ is much larger than $a^{-2}$ and, at low temperature, we use Eq. (\ref{EC2}) to 
calculate the Casimir energy per unit area to find ${E_C \over L^2}= -1.08\times 10^8$ eV m$^{-2}$. We obtain the Casimir pressure  using Eq.
(\ref{P9}) with $T=0$, and find $P=-1.35\times 10^{-4}$ Pa. We compare these numbers to those of the electromagnetic
Casimir effect for parallel plates at the same plate distance $a=1$  $\mu$m, where we find that the Casimir energy per unit area 
is ${E_C \over L^2}= -2.70\times 10^9$ eV m$^{-2}$, and the Casimir pressure is $P=-1.30\times 10^{-3}$ Pa, 1 to 2 orders
of magnitude larger than what we obtain for the charged scalar field using Eqs. (\ref{EC2}) and  (\ref{P9}).


\end{document}